\documentclass[twocolumn,amssymb, nobibnotes, showpacs, superscriptaddress, aps, prd]{revtex4-1}
\usepackage{graphicx,amsmath,amssymb,color}

\begin{document}
\title{Breaking the energy-symmetry blockade in magneto-optical rotation}

\author{Chengjie Zhu}
\affiliation{MOE Key Laboratory of Advanced Micro-Structured Materials, School of Physics Science and Engineering, Tongji University, Shanghai, China 200092}
\author{Feng Zhou}
\affiliation{Engineering Physics Division, National Institute of Standards \& Technology, Gaithersburg, Maryland USA 20899}
\affiliation{Wuhan Institute of Physics \& Mathematics, Chinese Academy of Sciences, Wuhan, China 430071}
\affiliation{School of Optical and Electronic Information, Huazhong University of Science \& Technology, Wuhan China 430074}
\author{Eric Y. Zhu}
\affiliation{Quantum Measurement Division, National Institute of Standards \& Technology, Gaithersburg, Maryland USA 20899}
\affiliation{Department of Electrical and Computer Engineering, University of Toronto, Toronto, Ontario, M5S 3G4, Canada}
\author{E. W. Hagley}
\affiliation{Sensor Science Division, National Institute of Standards \& Technology, Gaithersburg, Maryland USA 20899}
\author{L. Deng}
\email[Corresponding author:]{lu.deng@nist.gov}
\affiliation{Quantum Measurement Division,National Institute of Standards \& Technology, Gaithersburg, Maryland USA 20899}

\date{\today}

\begin{abstract}
	The magneto-optical polarization rotation effect has prolific applications in various research areas spanning the scientific spectrum including space and interstellar research, nano-technology and material science, biomedical imaging, and sub-atomic particle research. In nonlinear magneto-optical rotation (NMOR), the intensity of a linearly-polarized probe field affects the rotation of its own polarization plane while propagating in a magnetized medium. However, typical NMOR signals of conventional single-beam $\Lambda-$scheme atomic magnetometers are peculiarly small, requiring sophisticated magnetic shielding under complex operational conditions. Here, we show the presence of an energy-symmetry blockade that undermines the NMOR effect in conventional single-beam $\Lambda-$scheme atomic magnetometers.  We further demonstrate, both experimentally and theoretically, an inelastic wave-mixing technique that breaks this NMOR blockade, resulting in more than five orders of magnitude ($>$300,000-fold) NMOR optical signal power spectral density enhancement never before seen with conventional single-beam $\Lambda-$scheme atomic magnetometers. This new technique, demonstrated with substantially reduced light intensities, may lead to many applications, especially in the field of bio-magnetism and high-resolution low-field magnetic imaging.
\end{abstract} 


\maketitle

	
Magneto-optical effects arise when the different polarized components of an electrical field resonantly coupling to magnetic transitions of an atom experience different magnetic dichriosm \cite{r1,r1b,r1c}. When the effects become dependent on the intensity of the electric field itself, the resulting rotation of the polarization plane of the field is referred to as nonlinear magneto-optical rotation (NMOR). All polarimetry-based alkali metal vapor atomic magnetometers employ this principle for extremely weak magnetic field detection. Using spin relaxation management techniques \cite{r2,r3,r4,r5,r6,r7,r8,r9,r10,r11}, together with state-of-the-art magnetic shieldings and phase-locking detection electronics, atomic magnetometers have demonstrated extremely high sensitivities \cite{r6,r8,r9,r10,r11,r12,r12a} approaching the benchmark performance of superconducting quantum interference devices \cite{r12b}. 

\begin{figure}[htb]
	\centering
	\includegraphics[width=8.5 cm]{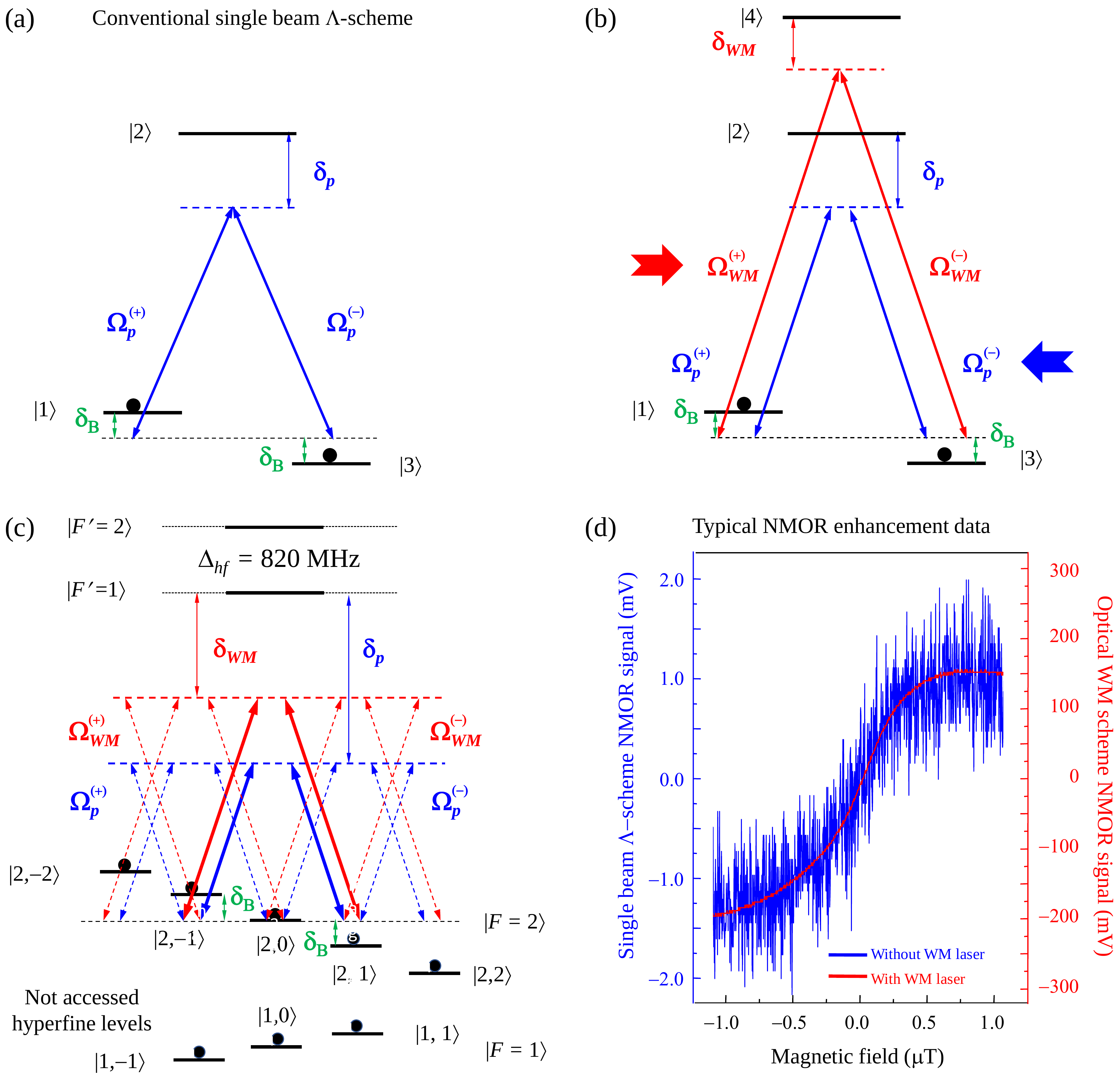}
	\caption{Simplified models, actual atomic system, and experimental observations. (a) Simplified $\Lambda-$scheme used in conventional single beam atomic magnetometers. The lower two states are the $m_F=\pm 1$ hyperfine states of a generic $F=1$ manifold (the $m_F=0$ state is not shown). $\Omega_p^{(\pm)}$ represents $\sigma^{(\pm)}$ components of a linearly-polarized probe field. (b) Simplified wave mixing scheme where the inelastic scattering from the wave-mixing (WM) field to the probe field is the underlying physical process. $\Omega_{\rm WM}^{(\pm)}$ represents $\sigma^{(\pm)}$ components of the linearly-polarized wave-mixing (WM) field. (c) Actual experimental system (magnetic sub-levels in the excited states are omitted) where the thick arrows depict the part that corresponds to (b). (d) Typical experimental data showing a 150-fold NMOR optical SNR enhancement (see Supplementary Information). }\label{fig:fig1}
\end{figure}

The core element of polarimetry-based alkali vapor atomic magnetometers is a three-state atomic system [Fig.~\ref{fig:fig1}(a)] 
coupling to a linearly polarized probe field \cite{r13,r14,r15,r15a,r15b,r15c}. Although the NMOR signal is enhanced by the probe intensity in this simple $\Lambda$-scheme, the NMOR optical signal-to-noise ratio (SNR) is generally low, requiring complex magnetic shielding and sophisticated electronics in addition to long data-acquisition times. Here, we introduce an optical inelastic wave-mixing (WM) process via a second light field counter-propagating with respect to the probe field.  This inelastic WM process opens a multi-excitation channel by sharing two fully-occupied intermediate states \cite{r17,r17a,r18} with the probe field. Consequently, it significantly modifies the ground-state Zeeman coherence and the magnetic dichroism at the probe frequency. The key underlying physics of this new technique is the removal of the energy-symmetry-based NMOR blockade inherent in conventional single-beam $\Lambda-$type atomic magnetometers, unleashing the full propagation strength of the NMOR effect and giving rise to more than five orders of magnitude NMOR optical signal power spectral density SNR enhancement. 

\vskip 10pt
\noindent{\bf Theoretical framework.} Consider the simplified four-state atomic system, depicted in Fig.~\ref{fig:fig1}(b), where atomic state $|j\rangle$ has energy $\hbar\omega_j$ ($j=1,...,4$). We assume that the probe field $\mathbf{E}_p$ (frequency $\omega_p$) is polarized along the $\hat{x}$-axis and propagates along the $\hat{z}$-axis. Its $\sigma^{(\pm)}$ components couple independently the $|1\rangle\leftrightarrow|2\rangle$ and $|3\rangle\leftrightarrow|2\rangle$ transitions with a large one-photon detuning $\delta_p=\delta_2=\omega_p-[\omega_2-(\omega_1+\omega_3)/2]$, forming two-photon transitions between states  $|1\rangle\leftrightarrow|3\rangle$ with a two-photon detuning $2\delta_B$. Here, the Zeeman frequency shift $\delta_B=g\mu_0 B$ in the axial magnetic field $B=B_z$ referenced to the mid-point, i.e., the $m_F=0$ state, between the two equally but oppositely shifted Zeeman levels $|1\rangle=|-1\rangle$ and $|3\rangle=|+1\rangle$ (see Figs. 1a and 1b). An optical WM field $\mathbf{E}_{\rm WM}$ (frequency $\omega_{\rm WM}$) propagates along the $-\hat{z}$-axis but has its linear polarization making an angle $\theta$ with respect to the $\hat{x}$-axis. Its
$\sigma^{(\pm)}$ components independently couple the $|1\rangle\leftrightarrow|4\rangle$ and $|3\rangle\leftrightarrow|4\rangle$ transitions with a large one-photon detuning $\delta_{\rm WM}=\delta_4=\omega_{\rm WM}-[\omega_4-(\omega_1+\omega_3)/2]$, forming two-photon transitions between states  $|1\rangle\leftrightarrow|3\rangle$ with the same two-photon detuning $2\delta_B$. Therefore, we have two $\Lambda$ transitions that share the same equally-populated states $|1\rangle$ and $|3\rangle$, creating a mutually influencing ground-state Zeeman coherence. We note that this is very different from the conventional double-Lambda four wave mixing process~\cite{r18,r18a,r18b,r18c,r18d} where the intermediate state has negligible Zeeman coherence [see discussion on Fig. S3].


The total electric field is given as $\mathbf{E}=\mathbf{E}_p+\mathbf{E}_{\rm WM}$ where in general, $\mathbf{E}_p=\left(\mathbf{\hat{e}}_x{\cal E}^{\rm (p)}_x+\mathbf{\hat{e}}_y{\cal E}^{\rm (p)}_y\right){\rm e}^{i\theta_p}+{\rm c.c.}$ and $\mathbf{E}_{\rm WM}=\left(\mathbf{\hat{e}}_x{\cal E}^{({\rm WM})}_x+\mathbf{\hat{e}}_y{\cal E}^{({\rm WM})}_y\right){\rm e}^{i\theta_{\rm WM}}+{\rm c.c.}$. Here, $\theta_{\rm p}=k_{\rm p}z-\omega_{\rm p}t$, $\theta_{\rm WM}=-k_{\rm WM}z-\omega_{\rm WM}t$, and $k_{\rm p(WM)}=\omega_{\rm p(WM)}/c$ is the wavevector of the probe (WM) field.  Expressing the electric field in a rotating basis using the relation $\mathbf{\hat{e}}_x=(\mathbf{\hat{e}}_{+}+\mathbf{\hat{e}}_{-})/\sqrt{2}$ and $\mathbf{\hat{e}}_y=-i(\mathbf{\hat{e}}_{+}-\mathbf{\hat{e}}_{-})/\sqrt{2}$ with $\mathbf{\hat{e}}_\pm$ being the unit vectors in the rotation bases, we then have  $\mathbf{E}_p=(\mathbf{\hat{e}}_+{\cal E}_p^{(+)}+\mathbf{\hat{e}}_-{\cal E}_p^{(-)}){\rm e}^{i\theta_p}+{\rm c.c.}$ with ${\cal E}_p^{(\pm)}=({\cal E}^{(\rm p)}_x\pm i{\cal E}^{(\rm p)}_y)/\sqrt{2}$  and $\mathbf{E}_{\rm WM}=(\mathbf{\hat{e}}_+{\cal E}_{\rm WM}^{(+)}+\mathbf{\hat{e}}_-{\cal E}_{\rm WM}^{(-)}){\rm e}^{i\theta_{\rm WM}}+{\rm c.c.}$ with ${\cal E}_{\rm WM}^{(\pm)}=({\cal E}^{(\rm WM)}_x\pm i{\cal E}^{(\rm WM)}_y)/\sqrt{2}$. 

Under the electric-dipole and the rotating-wave approximations, the interaction Hamiltonian of the system in the interaction picture can be expressed as
\begin{eqnarray}
\hat{H}\!=\!\sum_{j=1}^4\hbar\delta_j|j\rangle\langle j|\!+\!\hbar\!\sum_{m=2,4}\!\left[\Omega_{m1}|m\rangle\langle 1|\!+\!\Omega_{m3}|m\rangle\langle 3|\!+\!{\rm c.c}\right],\;\;
\end{eqnarray}
%
%
%
%
where $\delta_j$ is the laser detuning from state $|j\rangle$. Here, the half-Rabi frequencies of the probe field are  $\Omega_{21}=D_{21}{\cal E}_p^{(+)}/(2\hbar)$ and $\Omega_{23}=D_{23}{\cal E}_p^{(-)}/(2\hbar)$. The half-Rabi frequencies of the WM field are $\Omega_{41}=D_{41}{\cal E}_{\rm WM}^{(-)}/(2\hbar)$ and $\Omega_{43}=D_{43}{\cal E}_{\rm WM}^{(+)}/(2\hbar)$ because it counter-propagates with respect to the magnetic field direction. $D_{ij}=\langle i|\hat{D}|j\rangle$ is the transition matrix element of the dipole operator $\hat{D}$.

The dynamic behavior of the atomic system can be obtained by solving the Liouville equation given by
\begin{equation}\label{eq:Master}
i\hbar\frac{d}{dt}\hat{\rho}=\left[\hat{H},\hat{\rho}\right]-i\hbar\frac{1}{2}\left(\hat{\Gamma}\hat{\rho}+\hat{\rho}\hat{\Gamma}\right),
\end{equation}
where $\hat{\rho}$ is the density matrix operator of the atomic system.  $\hat{\Gamma}=\sum_{j=1,3}\gamma_{31}|j\rangle\langle j|+\Gamma_2|2\rangle\langle 2|+\Gamma_4|4\rangle\langle 4|$ with $\gamma_{13}=\gamma_{31}$ and $\Gamma_s\ (s=2,4)$ being the collisional (the Zeeman ground states) and radiative (the excited electronic states) decay rates, respectively.

Under the slowly varying envelope approximation, the Maxwell equations describing the evolution of the probe and WM fields are given as $(m=1,3)$
\begin{subequations}\label{eq:Max}
	\begin{align}
	& \left(\frac{1}{c}\frac{\partial}{\partial t}+\frac{\partial}{\partial z}\right){\cal E}^{(\pm)}_p=\kappa_{p}D_{m2}\rho_{2m},\\
	%
	%
	& \left(\frac{1}{c}\frac{\partial}{\partial t}-\frac{\partial}{\partial z}\right){\cal E}^{(\mp)}_{\rm WM}=\kappa_{\rm WM}D_{m4}\rho_{4m},
	%
	%
	\end{align}
\end{subequations}
where $\kappa_{p\,(\rm WM)}={\cal N}_a\omega_{p\,(\rm WM)}/(\epsilon_0c)$ with ${\cal N}_a$, $\epsilon_0$, and $c$ being the atom number density, vacuum permittivity, and speed of light in vacuum, respectively. Equations (1-3), which include 16 equations of motion for density matrix elements and 4 Maxwell equations for all circularly-polarized components of the two electrical fields, can be solved numerically to yield the probe-field dynamics from which all magneto-optical parameters can be evaluated.

\vskip 10pt
\noindent{\bf Enhancement of NMOR optical SNR: Physical intuition.}  Intuitive understanding of the large NMOR optical SNR enhancement and Zeeman coherence evolution can be gained by considering the steady-state solutions of Eq. (2) in the thin-medium limit. This approximation applies in the region near the entrance where changes in the probe field are not significant and a perturbation treatment for the NMOR signal is valid (see Methods). To this end, we assume that initially all atoms have their population shared equally between the states $|1\rangle$ and $|3\rangle$, but there is no appreciable initial Zeeman coherence \cite{r19}. 

In the thin-medium limit, the probe NMOR effect can be expressed perturbatively as [see Eq. (5) in Methods],
\begin{equation}
\rho_{21}^{(3)}-\rho_{23}^{(3)}={\cal R}_{\Lambda}+{\cal R}_{\rm WM},\nonumber
\end{equation}
where the superscript indicates the perturbation order and 
\begin{align} 
& {\cal R}_{\Lambda}\approx\frac{2i\gamma_{21}\rho_{11}^{(0)}}{\delta_2^2(\delta_2+i\gamma_{21})}\left[\frac{|\Omega_{23}^{(1)}|^2\Omega_{21}^{(1)}}{(2\delta_B+i\gamma_{31})}+\frac{|\Omega_{21}^{(1)}|^2\Omega_{23}^{(1)}}{(2\delta_B-i\gamma_{31})}\right],\nonumber\\
& {\cal R}_{\rm WM}\approx\frac{2i\gamma_{41}\rho_{11}^{(0)}}{\delta_4^2(\delta_2+i\gamma_{21})}\left[\frac{\Omega_{34}^{(1)}\Omega_{41}^{(1)}\Omega_{23}^{(1)}}{(2\delta_B+i\gamma_{31})}+\frac{\Omega_{14}^{(1)}\Omega_{43}^{(1)}\Omega_{21}^{(1)}}{(2\delta_B-i\gamma_{31})}\right].\nonumber
	%
\end{align}
Here, ${\cal R}_{\Lambda}$ is the NMOR of the conventional single-beam $\Lambda-$scheme [blue arrows in Fig.~\ref{fig:fig1}(b)] and ${\cal R}_{\rm WM}$ is the contribution to the total probe NMOR arising from the WM field [red arrows in Fig.~\ref{fig:fig1}(b)]. This WM contribution has more profound physical consequences than just a simple additional polarization source term for the electrical fields. First, notice that the terms in the square bracket in ${\cal R}_{\rm WM}$ have the identical Zeeman frequency shift denominators as that in ${\cal R}_{\Lambda}$, and this implies it has an identical magnetic resonance line shape, a critical magnetic-field-sensitivity-preserving feature of the optical WM method.

From a nonlinear optics point of view, because the Zeeman states $|1\rangle$ and $|3\rangle$ are initially equally populated there are two competing two-photon processes that drive the population transfer and establish the ground-state Zeeman coherence in the conventional single-beam $\Lambda-$scheme [Fig.~\ref{fig:fig1}(a)]. The two-photon transition $|1\rangle\rightarrow|3\rangle$ requires the absorption of $\Omega_p^{(+)}$ photons followed by emission of $\Omega_p^{(-)}$ photons and the two-photon transition $|3\rangle\rightarrow|1\rangle$ requires absorption of $\Omega_p^{(-)}$ photons followed by emission of $\Omega_p^{(+)}$ photons. Since the probe field is the only source of energy, the symmetric nature of these two two-photon processes dictates that both probe polarization components cannot change appreciably, resulting in \cite{r19a}
\begin{equation}
\frac{\partial|\Omega_p^{(\pm)}(z)|}{\partial z}\approx 0\;\;\Rightarrow{\rm NMOR\;\; angle}\;\;\alpha\propto {\cal N}_aL,\nonumber
\end{equation}
where $L$ is the medium length. This energy-symmetry blockade is the primary reason why the source term ${\cal R}_{\Lambda}$ for the probe field, having a third-order nonlinearity, results in a NMOR angle $\alpha$ rotation that is only linearly proportional to the atom density and medium length. It is this {\bf energy-symmetry blockade} that suppresses the full non-linear propagation strength of the NMOR effect in single-beam $\Lambda-$ type magnetometry. The WM field in the WM scheme (Fig. 1b) introduces an inelastic WM and scattering process (see Fig. S3 in Supplementary Material) that allows coherent energy transfer from the WM field to both the probe field and atomic medium, breaking the probe energy-symmetry blockade and enabling nonlinear growth of the probe polarization components.  The result is the unprecedented NMOR optical SNR enhancement [see Fig.~\ref{fig:fig1}(d)]. 

\begin{figure*}[htb]
	\centering
	\includegraphics[width=16 cm,angle=0]{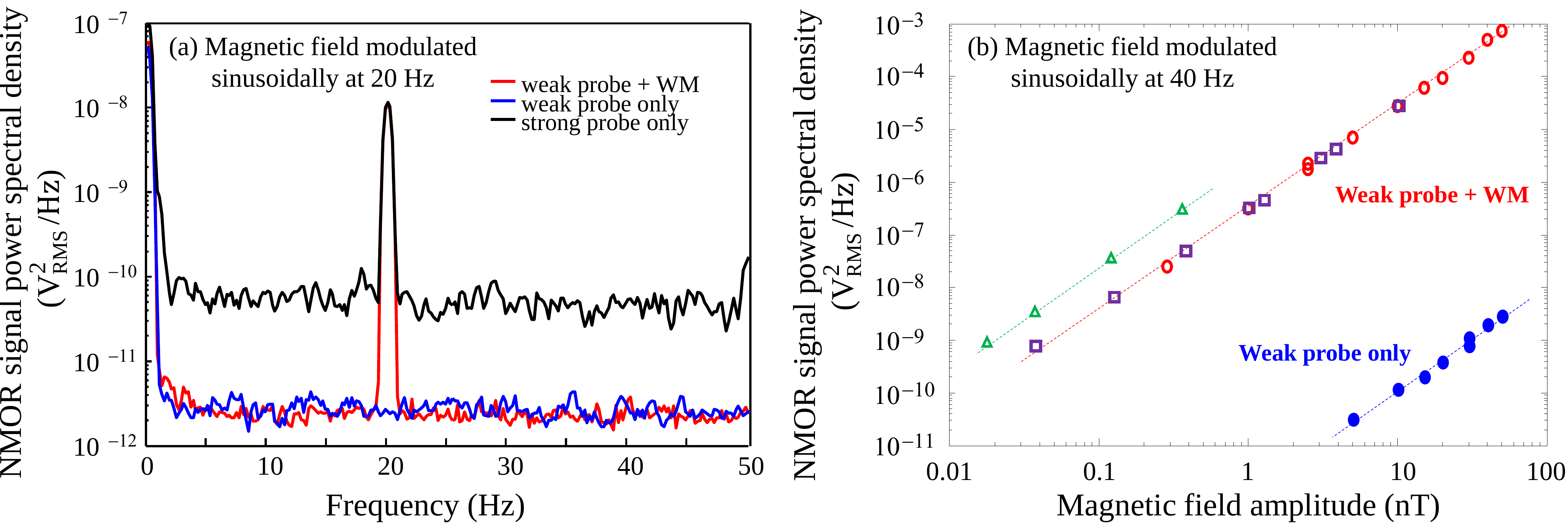}\\
	\caption{NMOR optical signal power spectral density and peak NMOR signal power spectral density vs. magnetic field amplitude at 311 K. (a) A 10 nT magnetic field is modulated sinusoidally at 20 Hz. Blue: single beam $\Lambda$ technique with a probe field intensity $I_p$ = 16.8 $\mu$W/cm$^2$. Red: optical WM technique with $I_p$ = 16.8 $\mu$W/cm$^2$ and a WM field intensity $I_{WM}$ = 15 $\mu$W/$cm^2$. Black: single beam $\Lambda$ technique with a probe field intensity $I_p$ = 2.6 mW/cm$^2$. (b) NMOR optical power spectral density as a function of magnetic field amplitude. Blue dots: single beam $\Lambda$ technique, $I_p$ = 650 $\mu$W/cm$^2$. Red open-circle and purple squares: optical WM technique with $I_p$ = 650 $\mu$W/cm$^2$ and $I_{WM}$ = 60 $\mu$W/cm$^2$. Green triangle: WM technique, but with $I_p$ = 1.3 mW/cm$^2$. Notice the more than 300,000-fold optical signal power spectral density enhancement achieved with the optical WM technique in both plots. }\label{fig:fig2}
\end{figure*}
Figure~\ref{fig:fig2} further demonstrates the exceptionally low noise performance, large dynamic range, and more than 300,000-fold NMOR optical signal power spectral density enhancement using the inelastic WM technique. In Fig.~\ref{fig:fig2}(a), we show the NMOR optical signal power spectral density using a magnetic field of 10 nT modulated sinusoidally at 20 Hz. We first chose a very weak probe field so that no NMOR signal can be observed when the WM field was absent (blue trace). When the WM field was turned on a huge NMOR signal was observed (red peak).  However, the noise in the red trace is at the same level as that of the blue trace, indicating no appreciable additional noise was added, even though more than a 3 order of magnitude NMOR signal optical power spectral density enhancement was achieved. We then switched off the WM field and increased the probe intensity to match the NMOR signal amplitude set by the WM field.  This single-beam $\Lambda$ technique requires a substantially higher probe field intensity and as a result it exhibits a nearly two order of magnitude higher noise level (Fig.~\ref{fig:fig2}(a), black traces).  Figure~\ref{fig:fig2}(b) illustrates an even more astonishing NMOR optical power spectral density enhancement with a large magneto-optical dynamic range response. Here, we plot the NMOR signal peak value as a function of the magnetic field amplitude. The greater-than 300,000-fold optical signal power spectral density enhancement over the conventional single-beam $\Lambda$ technique demonstrates the superior performance and robustness of the optical WM technique. We emphasize that the data shown in Fig.~\ref{fig:fig2}b exhibit perfect linearity with an identical slope for both the optical WM technique and the single beam $\Lambda-$technique. This indicates that the optical WM technique can reach the same detection sensitivity of current state-of-the-art single beam $\Lambda-$transition-based technologies under the same near zero-field conditions at ambient temperature, but with a more than 10$^5-$fold enhancement in the NMOR optical signal power spectral density. 

Another interesting feature introduced by the WM field can also be argued intuitively. When the linear polarization of the WM field makes an angle $\theta$ with respect to the linear polarization of the probe field, its Rabi frequencies acquire $\theta-$dependence in the $\hat{\bf e}_{\pm}$ basis because of the projection of the WM field. As a result, $\Omega_{41}$ and $\Omega_{43}$ in ${\cal R}_{\rm WM}$ together contribute a factor of sin(2$\theta$) in ${\cal R}_{\rm WM}$. Therefore, the contribution ${\cal R}_{\rm WM}$ is maximized at the cross-polarization angle $\theta=\pi/4$ \cite{r1b}. At $\theta=0$ and $\pi/2$, ${\cal R}_{\rm WM}=0$ and the NMOR effect is negligible precisely because of the probe energy-symmetry NMOR blockade described above.   

\vskip 10 pt
\noindent{\bf Enhancement of NMOR optical SNR: Numerical calculations.} To obtain full probe-field polarization dynamics, we numerically integrated 16 atomic density matrix equations and 4 Maxwell equations (see Methods for all calculation parameters). 
\begin{figure*}[htb]
	\centering
	\includegraphics[width=16 cm]{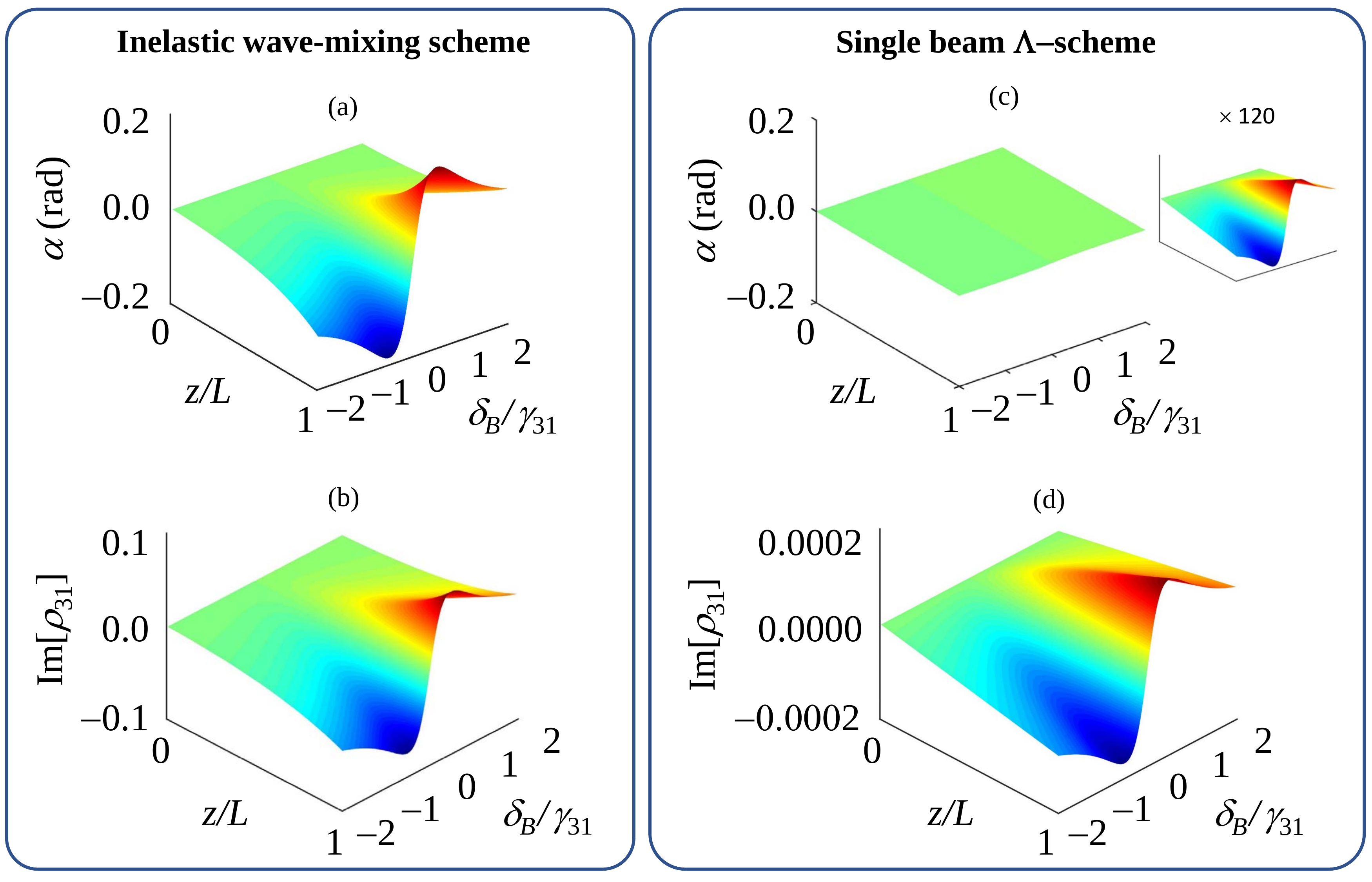}
	\caption{Probe polarization rotation and ground state Zeeman coherence: Probe polarization rotation $\alpha$ and the imaginary part of the ground state Zeeman coherence Im[$\rho_{31}$] as functions of the normalized Zeeman shift $\delta_B/\gamma_{31}$ and propagation distance $z/L$. The inset in the upper-right plot shows the actual line shapes of the magnetic resonance of the conventional single-beam $\Lambda$ method after  120-fold enlargement of the vertical axis. Notice that the resonance line shapes are identical to that of the optical WM method (upper-left panel), indicating the magnetic-field-sensitivity-preserving feature of the optical WM method. The imaginary part of the ground state Zeeman coherence $\rho_{31}$ exhibits nonlinear(left)/linear(right) growth characteristics in the WM/single-beam scheme, indicating large energy transfer in a deep-inelastic scattering process. }\label{fig3:PR}
\end{figure*}
In Fig.~\ref{fig3:PR} we show the probe NMOR angle $\alpha$ and the ground-state Zeeman coherence Im[$\rho_{31}$] which is responsible to the growth of $\alpha$ as functions of the normalized Zeeman frequency $\delta_B/\gamma_{31}$ and propagating distance $z/L$. Here, we have taken $\theta=\pi/4$ as discussed above [the effect of this angle is discussed in Fig. 4(a) and 4(b)]. Two of the most noticeable features shown in Fig.~\ref{fig3:PR} are: (1) the NMOR rotation by the optical WM technique [Fig.~\ref{fig3:PR}(a)] is more than 100 times larger than that of the conventional single beam $\Lambda-$technique [Fig.~\ref{fig3:PR}(c)], confirming our experimental observations shown in Fig.~\ref{fig:fig1}(d); and (2) the NMOR line shape of the optical WM technique is identical to that of the single beam $\Lambda-$technique [see the inset in Fig.~\ref{fig3:PR}(c)] which also agrees with experimental observations [see Fig.~\ref{fig:fig1}(d)]. This latter feature is critically important in applications since the resonance line shape determines the detection resolution and sensitivity for magnetic field detection. 

The ground-state Zeeman coherence is of central importance to magneto-optical effects arising from ground-state magnetic dichroism.  In the conventional single beam $\Lambda-$scheme, the NMOR-blockade restricts the growth of ground-state Zeeman coherence, limiting the growth of the NMOR effect [Fig.~\ref{fig3:PR}(d)]. The removal of this NMOR propagation blockade by the optical WM method enables significant ground-state Zeeman coherence growth [Fig.~\ref{fig3:PR}(b)] by efficient inelastic scattering (see discussion of Fig. S3 in Supplementary Material), resulting in a highly efficient NMOR optical SNR enhancement.  The significant difference in Im[$\rho_{31}$] attests to the fundamental difference between the two schemes. Indeed, it is quite astonishing that such a weak, symmetry-breaking WM field can lead to such a significant increase in the system's Zeeman coherence.

In the calculations above we chose the cross-polarization angle between the two fields as $\theta=\pi/4$. In the inelastic optical WM method the probe polarization rotation $\alpha$ is strongly dependent on this angle. 
\begin{figure*}[htb]
	\centering
	\includegraphics[width=16 cm]{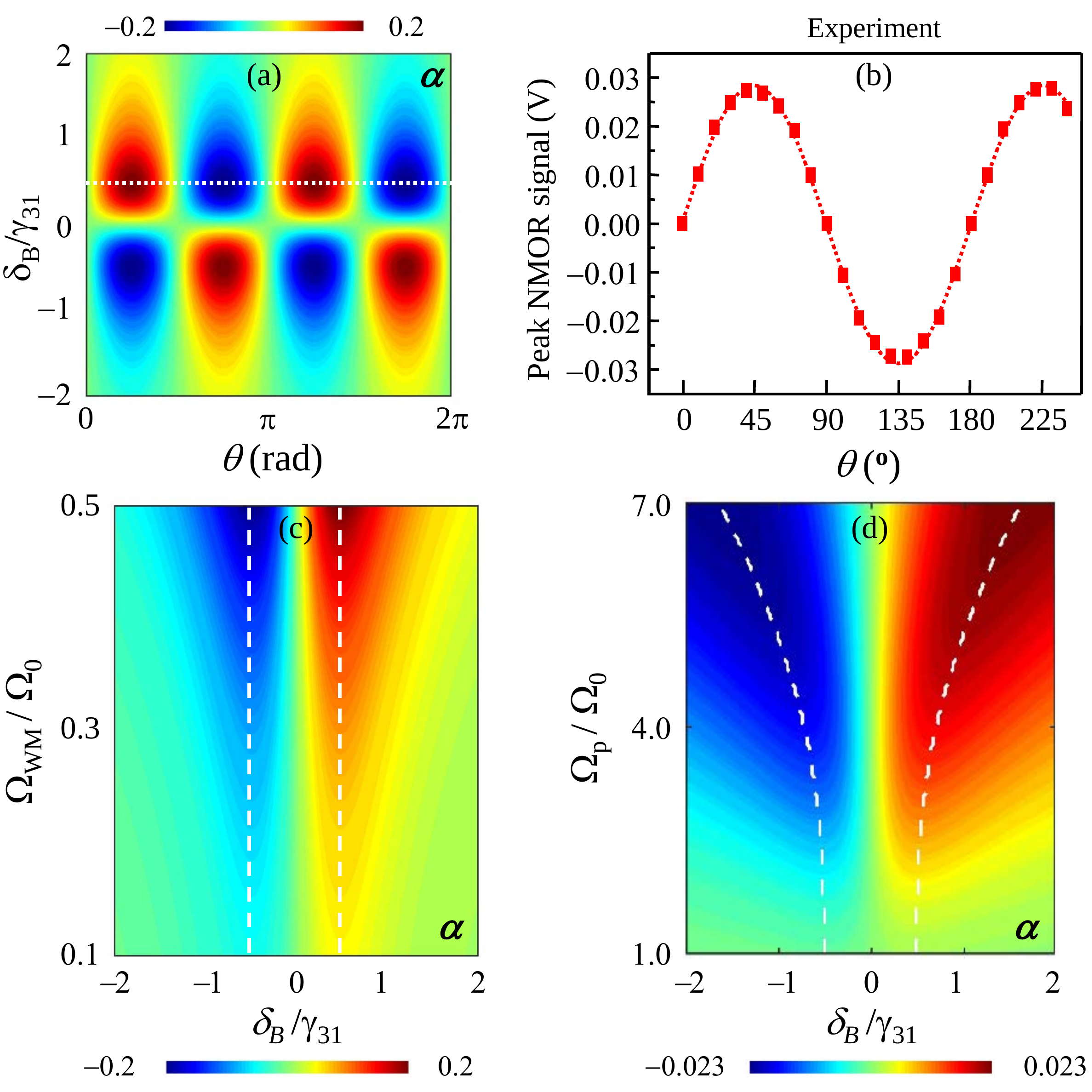}
	\caption{Polarization angular dependence of NMOR and the magnetic resonance linewidth preserving effect. (a) Calculated probe field NMOR angle $\alpha$ as a function of the normalized Zeeman frequency shift $\delta_B/\gamma_{31}$ and the angle $\theta$ between the polarizations of the probe and WM fields  at $Z/L=1$. The white dotted-line indicates the sinusoidal oscillation at $\delta_B/\gamma_{31}=0.5$. (b) The experimental NMOR peak signal (solid squares) plotted as a function of the angle $\theta$ between the polarization directions of the WM and the probe field. The axial magnetic field is 10 nT.  The red dotted line is a sine curve fit to the data. (c) and (d): Calculated power broadening of the magnetic resonance for the optical WM method (c) and the conventional single-beam $\Lambda$ method (d) as a function of the normalized WM and probe Rabi frequencies and the Zeeman shift (here, Rabi frequencies are normalized with respect to $\Omega_0=2\pi\times$ 200 kHz). White dashed lines indicate the peaks of the NMOR signal. No magnetic resonance broadening can be seen in the WM method, (c) whereas the power broadening of the NMOR signal is significant in the conventional single-beam $\Lambda$ method (d) even though the NMOR signal amplitude is still substantially less than that in (c), see color scales). Parameters are the same as in Fig. 2.}\label{fig4:theta}
\end{figure*}
In Fig.~\ref{fig4:theta}(a) we show the probe NMOR $\alpha$ as a function of the $\theta$ and $\delta_B/\gamma_{31}$.  Experimentally, we set a constant magnetic field and varied the angle between the linear polarizations of the two fields.  When this cross-polarization angle $\theta=\pi/4$, the NMOR optical SNR enhancement is maximized [Fig. 4(b)], in a full agreement with numerical calculations.

One of the advantages of the inelastic optical WM method is that the large NMOR optical SNR enhancement is achieved simultaneously with substantially reduced probe field intensities (the WM field intensity is even lower). This has a significant technological advantage in applications. In the conventional single-beam $\Lambda$ method, the probe intensity must be sufficiently high to bring out the NMOR effect (typically, more than 500 $\mu$W/cm$^2$ probe intensities are used with substantially smaller one-photon detunings).  Using the WM method, we observed NMOR signals with only a probe intensity of 10 $\mu$W/cm$^2$, and with a one-photon detuning 10 times larger. Under these light-level conditions the conventional single-beam $\Lambda-$scheme yields no detectable signal, even with long data acquisition times. One immediate benefit of such a low-level optical power operation is the possibility of achieving an unprecedented magnetic field detection sensitivity using atomic species with Zeeman de-coherence rates of less than 1 Hz.  Indeed, under the typical light intensity level, single-beam $\Lambda$ methods will introduce an intolerable power broadening \cite{r20,r21} that degrades the magnetic resonance line shape of ground-state Zeeman levels with such small de-coherence rates. 

To demonstrate the advantage of the low light level NMOR effect using the optical WM method, we show in Fig. 4(c) the numerical calculation of the probe NMOR signal as a function of the normalized Zeeman shift frequency $\delta_B/\gamma_{31}$ and the WM field Rabi frequency $\Omega_{\rm WM}/\Omega_{0}$ (here we fix $\Omega_p(0)=\Omega_0=2\pi\times$ 200 kHz). The white dashed lines indicates the peaks of the NMOR signal, showing the preserved magnetic resonance width.  However, the calculated probe NMOR signal of the conventional single-beam $\Lambda-$scheme [Fig. 4(d)] exhibits substantial power broadening as the probe intensity increases. Notice that even with the probe field as strong as $\Omega_p=7\Omega_0=2\pi\times$1.4 MHz, where the magnetic resonance line width has doubled from power broadening, the corresponding NMOR signal strength is still nearly a factor of 10 smaller than that shown in Fig. 4(c) (see the color scales).  This demonstrates that the conventional single beam method will encounter a significant difficulty when applied to very low magnetic field detection using an atomic species with a ground-state Zeeman de-coherence rate of $\gamma_{31}<$ 1 Hz. 

We have shown, both experimentally and theoretically, an inelastic WM technique that removes the energy-symmetry-based NMOR blockade inherent in the widely-used conventional single-beam $\Lambda$ method and enables an unprecedented NMOR signal SNR. The principle and technique demonstrated in our work may also facilitate applications employing similar fundamental physical principles. 


\begin{acknowledgements}
	CJZ acknowledge the National Key Basic Research Special Foundation (Grant No. 2016YFA0302800); the 973 Program (Grant No. 2013CB632701); the Joint Fund of the National Natural Science Foundation (Grant No. U1330203).
\end{acknowledgements}

%


\begin{thebibliography}{99}
\bibitem{r14} There are other transitions in a warm vapor that may contribute to the NMOR enhancement effect even when the non-accessed hyperfine states have been optically pumped.  However, the simplified single beam $\Lambda-$scheme provides a simple but key understanding of the NMOR effect.
\bibitem{r17a} All pump-probe experiments, regardless of whether they employ saturated absorption spectroscopy or polarization spectroscopy, essentially rely on a two-level atomic system where both the pump and the probe are on resonance.  The pump field saturates the transition while the probe is scanned across the absorption profile. In all these operations only the linearized response of the probe field is considered. Our system does not satisfy any of these conditions.
\bibitem{r19}  This corresponds to thermally distributed populations when the two states are degenerate or nearly degenerate (neglecting vacuum induced spontaneous coherence).
\bibitem{r19a} In the thin medium limit all probe Rabi frequencies appear in ${\cal R}_{\Lambda}$, which is the polarization source for the probe Maxwell equations, are constant, i.e., $\Omega_{21}=\Omega_{23}=\Omega_{p}(0)/\sqrt{2}$.  This leads to $\alpha\propto {\cal N}_{a}L$  \cite{r15}.
\end{thebibliography}
\end{document}